\documentclass[useAMS,usenatbib,referee ]{mn2e}
\usepackage{amsmath,amssymb}
\usepackage{natbib}
\usepackage{graphicx}
\usepackage{times}
\usepackage{ulem}
\def\ifnote{\iffalse}
\newcommand{\lumfns}{luminosity function}
\newcommand{\lumf}{luminosity function~}
\def\vvmax{$\left\langle V/V_{\rm max}\right\rangle$}

\author[D. Wanderman \& T. Piran]{{David Wanderman$^1$ and Tsvi Piran$^2$}\\
Racah Institute of Physics, The Hebrew University, Jerusalem 91904, Israel.  $^1${email:david.wanderman@mail.huji.ac.il}
$^2${email:tsvi@phys.huji.ac.il}\\
}
\title{The luminosity function and the rate of {\it Swift}'s Gamma Ray Bursts}
\begin{document}
\date{\today}
\maketitle
\label{firstpage}

\begin{abstract}
We invert directly the redshift - luminosity distribution of observed long {\it Swift} GRBs to obtain their rate and \lumfns.
Our best fit rate is described by a broken power law that rises like $(1+z)^{2.1^{+0.5}_{-0.6}}$ for $0 < z < 3$ and decrease like $(1+z)^{-1.4^{+2.4}_{-1.0}}$ for $z > 3$.
The local rate is $\rho_0 \simeq 1.3^{+0.6}_{-0.7} [Gpc^{-3}yr^{-1}]$.
The luminosity function is well described by a broken power law with a break at $L* \simeq 10^{52.5\pm 0.2}[erg/sec]$ and with indices $\alpha = 0.2^{+0.2}_{-0.1}$ and $\beta = 1.4^{+0.3}_{-0.6}$. The recently detected GRB 090423, with redshift $\approx 8$, fits nicely into the model's prediction, verifying that we are allowed to extend our results to high redshifts. 
 While there is a possible agreement with the star formation rate (SFR) for $z < 3$, the high redshift slope is shallower than the steep decline in the SFR for $4 < z$. However we cannot rule out a GRB rate that follows one of the recent SFR models.

\end{abstract}

\section{Introduction}
Gamma-ray bursts (GRBs) are short and intense pulses of soft $\gamma$-rays.
In this work we study their \lumf and their cosmic rate.
These functions are essential to understand the nature of GRBs and to determine their progenitors.
They may also shed light on the still mysterious physics of the central engine. 

The \lumf and the cosmic GRB rate are observationally entangled as the observed rate is a convolution of the \lumf with the cosmic rate. For this reason, almost every work before this paper have made an a-priori assumptions on at least one of these functions.
At first - having no motivation to believe otherwise - the rate was assumed to be constant.
The simplest form for the \lumf was a standard candle i.e. a constant luminosity.
These early studies used the measured \vvmax~ value to find the typical luminosity \citep{Mao(1992),Piran(1992),Fenimore(1993),Ulmer(1995)a,Ulmer(1995)b} and correspondingly a maximal distance from which GRBs were observed.
Later, using the flux distribution ($logN-logP$ relation), \cite{Cohen(1995)} and \cite{Loredo(1998)} showed how relaxing the standard candle assumption allows a corresponding relaxation of the constant rate assumption.

\cite{Paczynski(1998)} noticed that the hosts of the bursts are in star forming regions and suggested that GRBs follow the SFR \citep[see also][]{Totani(1997),Wijers(1998)a}.
The detection of a supernova associated with GRB980425 \citep{Galama(1998)}, strengthened the expectation that the GRB rate should follow the SFR.
Using this proportionality, numerous studies examined the typical luminosity assuming at first standard candles \citep[e.g.][]{Wijers(1998)a,Totani(1999)} and later more elaborate shapes of the \lumf \citep[e.g.][]{Schmidt(1999),Schmidt(2001b),Schmidt(2001a),Guetta(2005),Firmani(2004),GuettaPiran(2005)}.

{\it Swift} that discovers routinely GRBs afterglows detected GRBs from higher redshifts than was previously possible sparked renewed interest in the GRB redshift distribution \citep[e.g.][]{Berger(2005),Natarajan(2005),Bromm(2006),Jakobsson(2006),Le(2007),Yuksel(2007),Salvaterra(2007),Liang(2007),Chary(2007)}.
More and more signs appeared suggesting that the rate of GRBs does not simply follow the global star formation rate, as was believed earlier.
\cite{Firmani(2005),LeFloch(2006),Daigne(2006),Le(2007),Guetta(2007)} conclude that the GRB rate differs from the SFR. In  particular we observe more high redshift bursts than what is expected for a rate that follows the SFR.
\cite{Yuksel(2008)} uses the high luminosity subsample of bursts to obtain the GRB rate without assuming any luminosity function. They find that the GRB rate at high redshifts ($z > 4$) is higher on than \cite{HB(2006)} SFR.
Alternatively luminosity evolution, has been suggested by some authors: \citep[e.g.][]{Lloyd-Ronning(2002),Firmani(2004),Matsubayashi(2006),Kocevski(2006)}.
\cite{Salvaterra(2008)} used a sample of long bursts with redshift and found evidence for luminosity evolution.
Other papers, including this one, find self consistency without a luminosity evolution.

We introduce here a new method for determining the rate and the luminosity, by inverting the observations without making any assumptions on the functional form of the luminosity and rate functions. This direct inversion allows us to use most of the available redshift data and to obtain robust estimation of both functions.

In \S\ref{sec:sample} we describe the sub-sample of bursts we analyze and its advantages over other samples. In \S\ref{sec:direct} we introduce a new formalism allowing us to directly invert the observed distribution and obtain the intrinsic \lumf and rate. In \S\ref{sec:results} we present the results obtained using this method and the confidence ranges associated with them. 
In \S\ref{sec:contosfr} we address the question whether the GRB rate is consistent with the SFR. We discuss a few implications of the results in \S\ref{sec:summary}.

\section{The Sample}
\label{sec:sample}
\begin{figure}
\includegraphics[width=391pt]{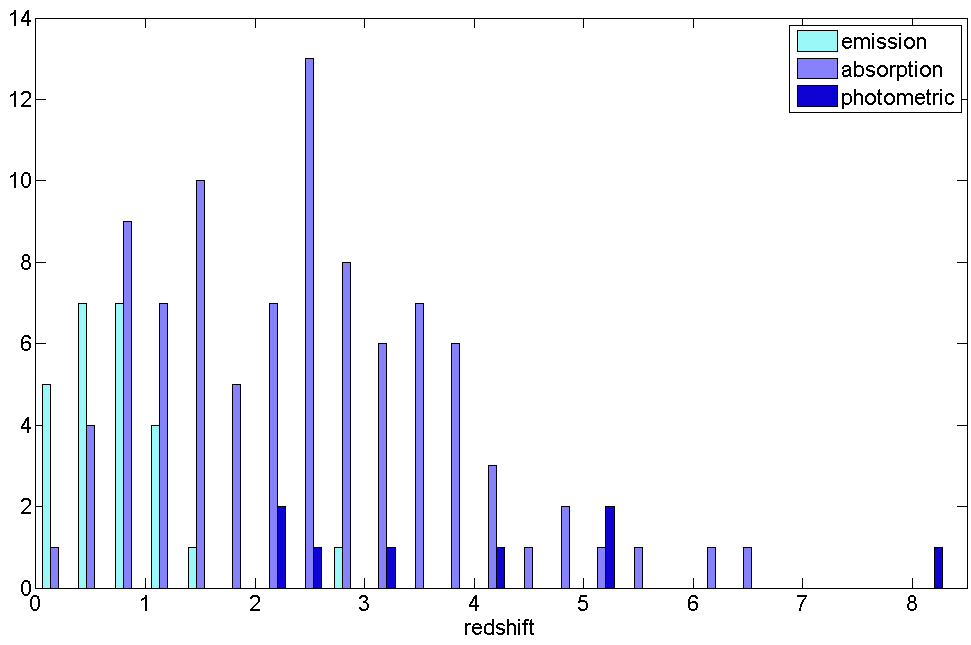}
\caption{The redshift distribution for the different methods: Absorption lines, Emission lines and Photometry}
\label{fig:zmethodsdist}
\end{figure}

We consider long bursts ($t_{90} \geq 2sec$) detected by {\it Swift} from the beginning of its operation until burst 090726
\footnote{All data were taken from the ${\it Swift}$ information page  ${http://swift.gsfc.nasa.gov/docs/swift/archive/grb\_table/}$}, with a measured peak-flux and a measured redshift.
Generally GRBs redshifts are obtained from the optical afterglow spectrum using absorption lines or photometry, or from the spectrum of the host galaxy using emission lines.
However in the global sample we find different redshift distributions for the different detection methods, namely: Absorption, Emission and Photometry (see Figure \ref{fig:zmethodsdist}).
For redshifts determined using the hosts' emission lines, we do not detect high-redshift events, whereas absorption lines redshifts extend over the entire range of redshifts. 
Furthermore, emission lines are more susceptible to a selection effect known as the 'redshift desert' in the range ${1.1 < z < 2.1}$ \footnote{Although a redshift determination through absorption lines is also difficult in the range ${1.5 < z < 2.1}$}, (\citealt{Fiore(2007)}, see also \citealt{Coward(2008)}).
This is in line with the fact that we have also found that the probability to measure the redshift using emission lines strongly depends on the gamma ray flux, favoring high fluxes
\footnote{A possible explanation is that a brighter burst can be more easily localized, making  follow-up  possible.}. This effect is mild for absorption lines and photometry (see appendix B).

To obtain a more uniform sample we consider therefore only bursts whose redshift was measured using the afterglow.
For each burst we calculate the isotropic equivalent peak-luminosity: $L_{iso}$ (see Appendix A. for details) using the peak-flux and redshift. We use standard $\Lambda CDM$ cosmology with $h = 0.7, \Omega_m = 0.27, \Omega_{\Lambda} = 0.73$.

\section{A Direct Estimate of the Luminosity Function and the Rate}
\label{sec:direct}
\subsection{Assumptions}
We assume that the \lumf is redshift independent. In this case the number of bursts at a given redshift and with a given luminosity is the product of the luminosity function, $\phi(L)$, that depends only on the luminosity $L$ and the GRB rate, $R_{GRB}(z)$, that depends only on the redshift $z$.
This common assumption is reasonable since a-priori there is no competing reason why the luminosity function should depend on the redshift.
We test later the validity of this assumption and show that it is accepted with a high statistical significance.

The isotropic peak luminosity function, $\phi(L)$, is defined traditionally as the fraction of GRBs with isotropic equivalent luminosities in the interval $\log L$ and $\log L + d\log L$.
The rate, $R_{GRB}(z)$, is defined as the co-moving space density of GRBs in the interval $z$ to $z + dz$.
The distribution density: $n(L,z)$ is given by:
\begin{equation} n(L,z)dlogLdz=\phi(L) \cdot R(z) dlogLdz \ , \end{equation}
where \begin{equation} R(z) = {\frac{R_{GRB}(z)}{(1+z)}} {\frac{dV(z)}{dz}} \ ,\end{equation}
is the differential co-moving rate of bursts at a redshift z, ${{dV(z)}/{dz}} $ is the derivative of the volume element and the factor $(1+z)^{-1}$ reflects the cosmological time dilation.

\subsection{Formalism}
We consider now a direct method to invert the observed $L - z$ distribution and obtain the functions $\phi$ and $R_{GRB}$. To do so we approximate $\phi(L)$ and $R(z)$ as $step \: functions$ whose range is divided to bins with a constant value within each bin. These functions can be expressed as a sum of Heaviside functions. 

First, we divide both luminosity and redshift intervals into bins.
We denote \begin{equation}\phi_i \equiv \phi(L_i \leq L < L_{i+1}) \ ,\end{equation}
 \begin{equation}
\label{eqn:rate_bins_decl}
R_j \equiv R(z_j \leq z < z_{j+1}) = 
\frac{1}{(z_{j+1} - z_j)} \int_{z_j}^{z_{j+1}}dz{\frac{R_{GRB}(z_j)}{(1+z)}}{\frac{dV(z)}{dz}}\ .\end{equation}
We also define the {\it weights factors}
\begin{equation} w_{ij} \equiv \int_{L_i}^{L_{i+1}}\int_{z_j}^{z_{j+1}} \theta(L,z)dlogLdz \ ,\end{equation}
 as the {\it probability for detecting a burst with a measured redshift z and luminosity L} 
where $\theta(L,z) \equiv \theta_z(p(L,z))$ is the probability to detect and measure redshift for a burst with a luminosity $L$ at a redshift $z$ (see appendix B). 
We denote the observed number of events per bin as:
\begin{equation} N_i \equiv N(L_i \leq L < L_{i+1}) \ , \end{equation}
\begin{equation} N_{,j} \equiv N(z_j \leq z < z_{j+1})	\ , \end{equation}
\begin{equation} N_{ij} \equiv N(L_i \leq L < L_{i+1} , z_j \leq z < z_{j+1}) \	, \end{equation}
\begin{equation} N \equiv \sum_{ij}N_{ij} \ . \end{equation}

Next, we determine $\phi_i$ and $R_j$ and the error estimates, using the maximum-likelihood formalism.
We define $M$ the (logarithmic) likelihood of the model given the observations as:
\begin{equation}
 M = \sum_{ij} N_{ij} \ln[ \phi_i R_j w_{ij}] - N \ln [\sum_{ij} \phi_i R_j w_{ij}] \ . \end{equation}
At the maximum all partial derivatives of $M$ with respect to $\phi_i$ and with respect to $R_j$ vanish,
leading to
\begin{equation}
 \label{eqn:rate1}
 \phi_i = \frac{N_i}{\sum_j R_j w_{ij}} {\frac{\sum_{i'j'} \phi_{i'} R_{j'} w_{i'j'}}{N}} \ , \end{equation}
 and
\begin{equation}  \label{eqn:lumf} R_j = {\frac{N_{,j}}{\sum_i \phi_i w_{ij}}} {\frac{\sum_{i'j'} \phi_{i'} R_{j'} w_{i'j'}}{N}} \ . \end{equation}
Notice that the second term on the right hand side of each of the equations \ref{eqn:rate1} and \ref{eqn:lumf} is the same normalization factor.
We have a set of non-linear equations with as many equations as variables.  We solve numerically these non linear equations using successive iterations until convergence. A-priori it is not clear whether $\phi$ and $R$ are uniquely determined and whether there is a solution at all. 
However, we find good convergence. We have examined a large set ($10^8$) of initial guesses where each component was randomly drawn from a uniform distribution.  We found a rapid convergence to a unique solution for all initial guesses, all reaching the requested accuracy of $10^{-6}$ with less than 25 iterations. We thus conclude that the existence of other stable solutions is very unlikely.

\subsection{Error Estimates}
We approximate the error as the value for which M deviate by -1 from it's maximum: 
(i.e. the likelihood is smaller by a factor $e$). This reflects an $1\sigma$ error for a normal-distribution.
\begin{equation} -1 = N_i \ln(1+\frac{\Delta\phi_i}{\phi_i}) - N \ln(1+\frac{N_i}{N}\frac{\Delta\phi_i}{\phi_i}) \ , \end{equation}
\begin{equation} -1 = N_{,j} \ln(1+\frac{\Delta R_j}{R_j}) - N \ln(1+\frac{N_{,j}}{N}\frac{\Delta R_j}{R_j}) \ . \end{equation}
The two solutions, i.e. the positive one and the negative one, give an upper and a lower bounds on the error respectively.
For small deviations we can approximate the error using the second derivatives of M:
\begin{equation} \frac{\Delta\phi_i}{\phi_i} \simeq \frac{\sqrt{2}}{\sqrt{N_i(1- N_i/N)}} \ , \end{equation}
\begin{equation} \frac{\Delta R_j}{R_j} \simeq \frac{\sqrt{2}}{\sqrt{N_{,j}(1- N_{,j}/N)}} \ . \end{equation}

To estimate the uncertainty induced by the specific bins choice, we preform all the analysis for a 1/2 unit redshift and $Log_{10}(L)$ binning and repeat for a 1/3 unit binning (where all bins widths are 1/3 unit, except the last two redshift bins which we cannot change because they contain too few data points). In the following, unless otherwise stated, we use the 1/2 unit binning for all further analysis. Clearly, the results  with different  binning are slightly different, but they are all within each other's error range.  When we include the uncertainty induced by the binning the error ranges become only slightly wider.

\section{Results}
\label{sec:results}
\begin{figure}
\includegraphics[width=470pt]{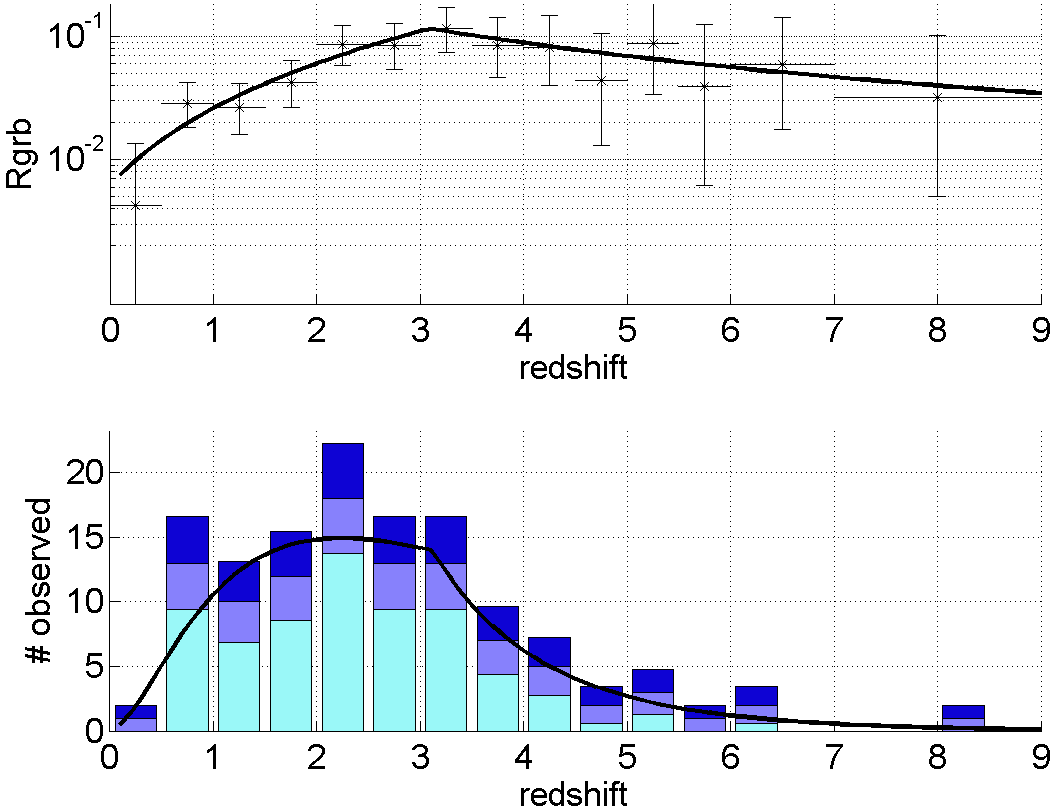}
\caption{The comoving rate and the observed events number redshift distributions}
Upper frame: The  resulting step function and the best fit model: A broken power law with indices $n_1 = 2.1$, $n_2 = -1.4$ and with a break at $z = 3.1$ (solid line). The 
$\chi^2$ values for the models is 2.3 at 10 d.o.f. giving a rejection probability of 0.007.
Lower frame: The number of detected bursts for each redshift bin and the rates expected from the model fitted above. The two upper boxes in each column represent the statistical error range.
\label{fig:rate}
\end{figure}

\begin{figure}
\includegraphics[width=470pt]{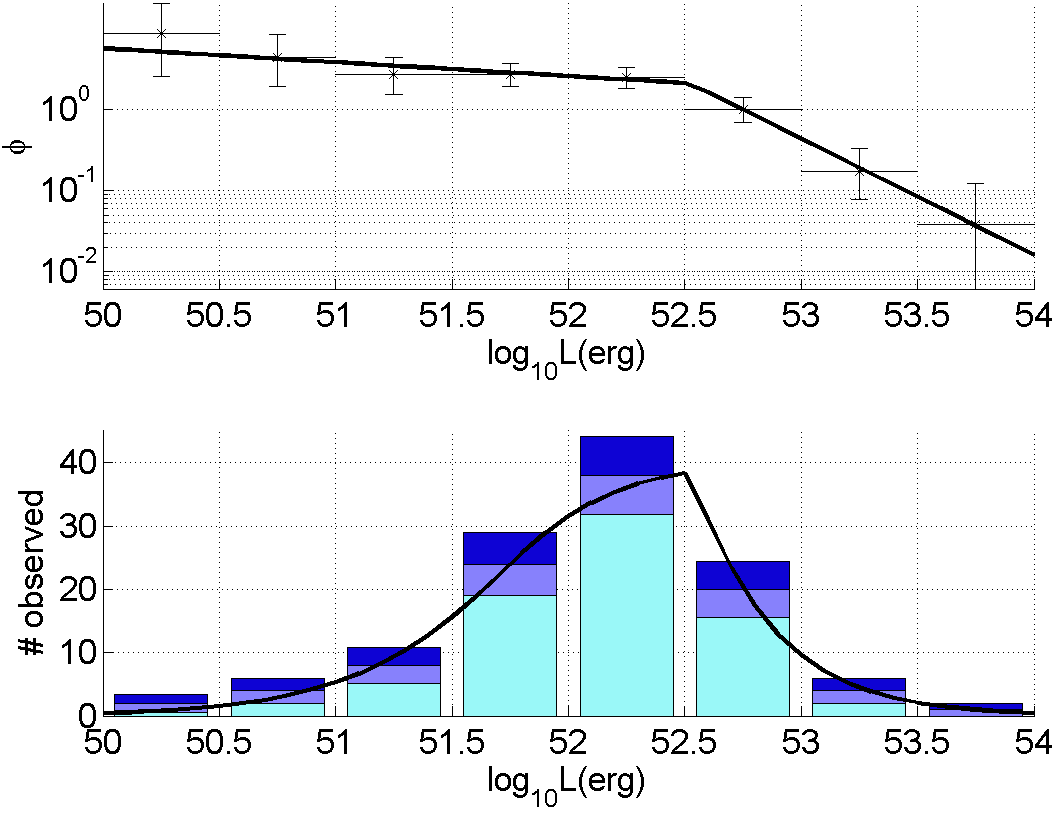}
\caption{The luminosity function and the observed and predicted luminosity distributions}
Upper frame: The resulting step function and the best fit model for the luminosity function: a broken power law with a break at $L*=10^{52.5}$, 
a low luminosity index $\alpha = 0.2$ and a high luminosity index $\beta=1.4$.
The $\chi^2$ value for this model is 0.63 at 4 d.o.f., giving a rejection probability of 0.04.
Lower frame: The number of detected bursts for each luminosity bin and the rates expected from the model fitted above. The two upper boxes in each column, represents the statistical error range.
\label{fig:luminosity}
\end{figure}

So far, we have not assumed any functional form for the luminosity function or for the rate as our method does not require  such an assumption. The ''raw" results are depicted in Figures \ref{fig:rate} and \ref{fig:luminosity}.  Later on we will compare these step functions with models for the GRB rate that follow the SFR.
However, in order to easily characterized the result we need to approximate our ''raw" step functions with simple functional forms. Therefore, after obtaining the results in the form of step functions  we approximate  them in term of  broken power laws:
 \begin{equation}
 \phi(L) =
\left\{
\begin{array}{ll}
(\frac{L}{L*})^{-\alpha}   & L < L* \ , \\
(\frac{L}{L*})^{-\beta} & L > L* \ . 
\end{array}
\right. 
\end{equation}
\begin{equation}
R_{GRB} = R_{GRB}(0) \cdot
\left\{
\begin{array}{ll}
  (1+z)^{n_1} & z \leq z_1 \ ,\\
   (1+z_1)^{n_1-n_2}  (1+z)^{n_2} & z > z_1 \ ,
\end{array}
\right. 
\end{equation}
We obtain the parameters of the best fit functions by minimizing the $\chi^2$ values.

The best fit functions are  shown (together with the step functions) in Figures \ref{fig:rate} and \ref{fig:luminosity}. 
The parameters of the best fit models are summarized in Table \ref{tbl:MC2}. 
To estimate the statistical errors involved in the parameter estimates we preformed a Monte-Carlo simulation. 
In this simulation we use the model with the best fit parameters to draw random sets of data (with same size as the real sample). We then carry out the analysis on this mock data sets and obtain new best fit parameters. Repeating this process many times we obtain a scatter of points in the parameters plane around the original best fit parameters. The central $68\%$ and $95\%$ ranges for each of the parameters separately, are also shown in Table \ref{tbl:MC2}.

\begin{table}\renewcommand{\arraystretch}{2}\addtolength{\tabcolsep}{-1pt}
\begin{center}
\begin{tabular}{|l||l|l|}
\hline
& $ 1/2 \: bins$ & $ 1/3 \: bins$  \\
\hline
$logL*$ & $52.53^{+0.24+0.54}_{-0.17-0.66}$ & $52.58^{+0.21+0.68}_{-0.14-0.25}$ \\
$\alpha$ & $0.17^{+0.19+0.37}_{-0.10-0.21}$ & $0.18^{+0.18+0.37}_{-0.09-0.20}$ \\
$\beta$ & $1.44^{+0.33+0.59}_{-0.64-7.08}$ & $1.59^{+0.36+0.66}_{-0.58-7.86}$ \\
$z_1$ & $3.11^{+0.63+1.35}_{-0.82-1.66}$ & $3.45^{+0.95+1.67}_{-0.83-2.64}$ \\
$n_1$ & $2.07^{+0.51+0.89}_{-0.63-1.55}$ & $1.74^{+0.41+0.83}_{-0.65-1.61}$ \\
$n_2$ & $-1.36^{+2.39+6.83}_{-1.00-1.41}$ & $-1.47^{+2.98+8.53}_{-1.30-1.66}$ \\
$\rho_0$ & $1.25^{+0.56+0.93}_{-0.70-1.52}$ & $1.71^{+0.76+1.27}_{-0.74-1.93}$ \\
\hline
\end{tabular}
\end{center}
\caption{Parameters results. The error ranges are 68\% and 95\% levels estimated using a Monte-Carlo simulations with 10000 sets, each of 101 data points.}
\label{tbl:MC2}
\end{table}

The luminosity function is well described by a broken power law, with a break at $L* \simeq 10^{52.5 \pm 0.2}[erg/sec]$ and with indices $\alpha = 0.2^{+0.2}_{-0.1}$ and $\beta = 1.4^{+0.3}_{-0.6}$. The broken power law fit is very good, giving $\chi^2 = 0.63$ for 4 d.o.f.
This result agrees with previous studies \citep[e.g.][]{Daigne(2006),Guetta(2007)}.
The \lumf cannot be fitted with a single power law, since such fit give high $\chi^2$, rejecting such a model with high significance (98\%). This contradict the results of \cite{Pelangeon(2008)} who studied the HETE-2 GRBs and found a consistency with a single power law \lumfns.
The rate is described as well by a broken power law for $1+z$, with a break at $z = 3.1^{+0.6}_{-0.8}$ and indices of $n_1 = 2.1^{+0.5}_{-0.6}$ and $n_2 = -1.4^{+2.4}_{-1.0}$.

The local event rate is $\rho_0 \simeq 1.3^{+0.6}_{-0.7} [Gpc^{-3}yr^{-1}]$, in agreement with previous studies e.g. \cite{Schmidt(1999)}[$\rho_0 \simeq 1.5$] (see however \cite{Schmidt(2001b)}[$\rho_0 \simeq 0.15$)]); \cite{Guetta(2005)}[$\rho_0 \simeq 0.5$]; \cite{GuettaDV(2007)}[$\rho_0 \simeq 1.1$]; \cite{Liang(2007)}[$\rho_0 \simeq 1.1$]; \cite{Pelangeon(2008)}[$\rho_0 \gtrsim 0.5$]. 
The main factors determining the low redshift (current) event rate are the overall GRB rate normalization, the low-redshift slope, the low end of the \lumf slope and most important the position of the low luminosity cutoff. This low luminosity cutoff is essential for any steep enough power law to prevent its divergence. This cutoff is 
critical to the question whether the model includes the low luminosity GRB population. The cutoff used in our estimates of  $\rho_0$ is  $L = 10^{50} erg/sec$.

\subsection{Consistency Checks}
\label{sec:consistency}
Before we can accept the model,we turn now to check the validity of the assumption  that the \lumf and the rate are independent. . To do so, we preform a two dimensional Kolmogorov-Smirnov test 
(see \cite{2DKS1}, \cite{2DKS2}, \cite{2DKS3}). 
In this test the two dimensional data is compared to the modeled distribution for each of 4 quadrant defined by axes crossing at each data point.
The maximal difference is used to estimate the probability that the data is drawn from the distribution implied by the model.
The probability that the models fits the data is displayed in the 2D K-S column in Table \ref{tbl:SFRsstats}.
The test give high probability ($96\%$), so we can accept the model and justify the underling assumption (of a redshift independent luminosity function).

\begin{figure}
\includegraphics[width=470pt]{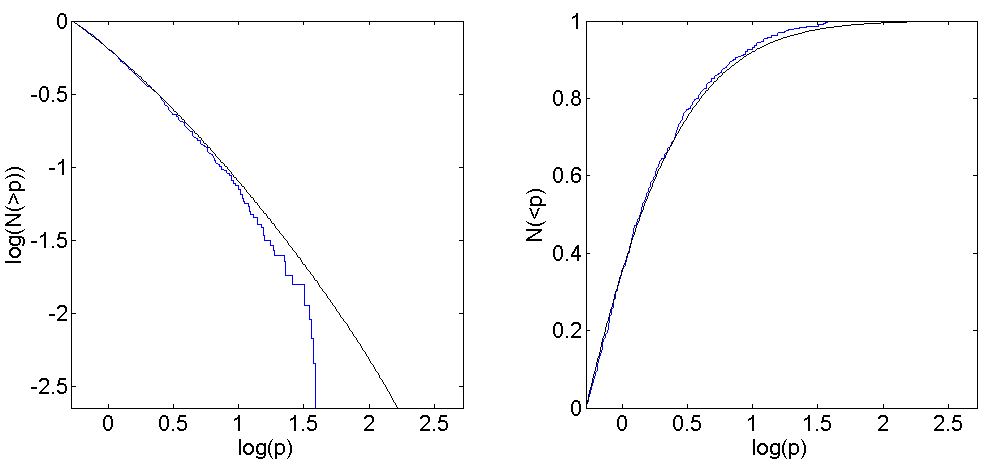}
\caption{Bursts count vs. peak flux for BATSE bursts}
Cumulative bursts distribution as a function of the peak flux $p$.
Left: logarithmic scale showing $N(<p)$    Right: log linear scale showing $N(>p)$, this is used for the KS-test, giving probability of $83\%$.
\label{fig:logNlogSBATSE}
\end{figure}

\begin{figure}
\includegraphics[width=470pt]{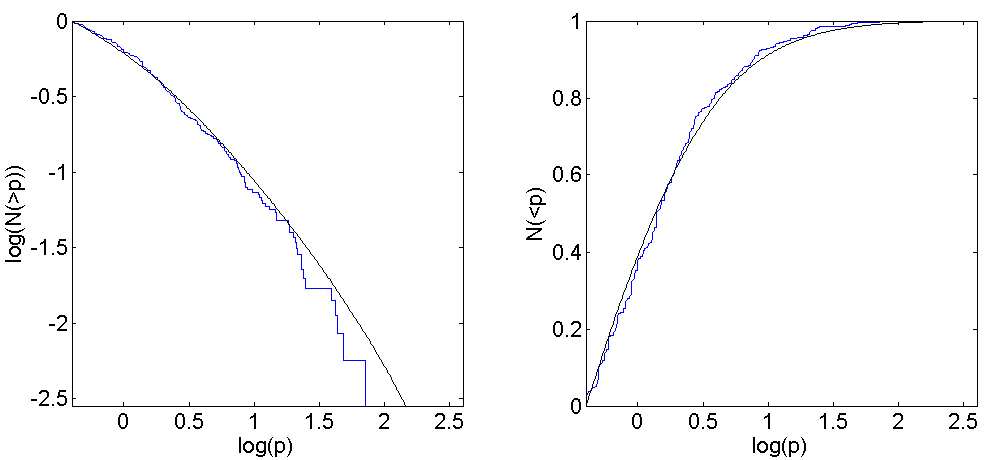}
\caption{Bursts count vs. peak flux for \it{Swift} bursts}
Cumulative bursts number distribution as a function of the peak flux $p$.
Left: logarithmic scale showing $N(<p)$    Right: log linear scale showing $N(>p)$, which is used for the KS-test, giving a probability of $17\%$.
\label{fig:logNlogSSwift}
\end{figure}

We carry out two other consistency checks.
First, we compare the peak flux distribution expected by our model with those observed by BATSE and by {\it Swift}.
The results are shown in Figures \ref{fig:logNlogSBATSE} and \ref{fig:logNlogSSwift}.
The KS-test results, 83\% for BATSE bursts and 17\% for the full sample of {\it Swift} bursts (with or without redshift), indicate acceptable model.
We note here, that the model is significantly rejected (KS-test $< 10^{-5}$) when compared to BATSE peak fluxes distribution using $P_{lim} = 0.25 ph/cm^2/sec$. However, when applying the effective detection threshold calculated by \cite{Band(2002)} of $P_{lim} = 0.525 ph/cm^2/sec$, the model is accepted with high significance (83\%).
Second, we compared the cumulative redshift distribution to the observed one and preformed a KS-test. Here, as well, the test  gives a high probability for accepting the model ($62\%$).

\begin{figure}
\includegraphics[width=470pt]{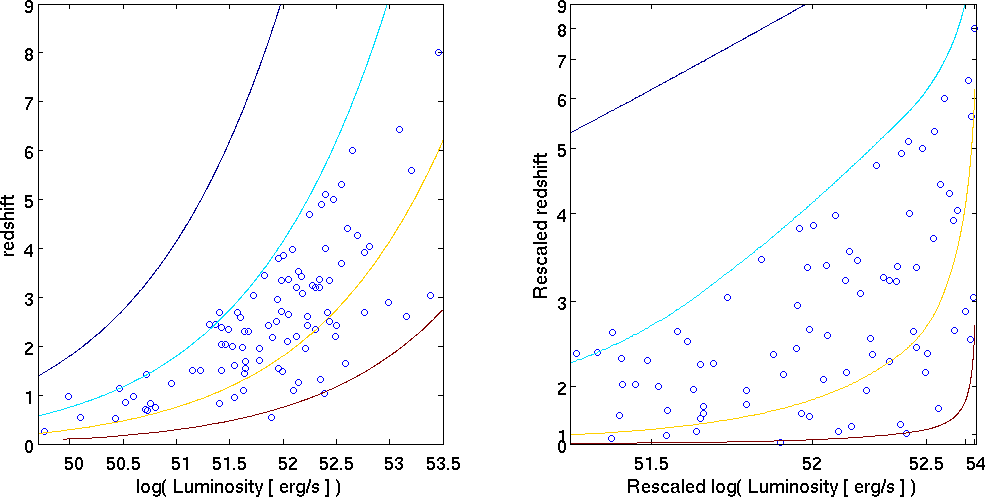}
\caption{Bursts on the L-z plane}
The bursts distribution in the Luminosity-Redshift plane.
Left: linear scale.    Right: Axes are rescaled using $R(z)$ and $\phi (L)$, so that we have a uniform bursts-number density.
The curved lines are contour lines for equal fluxes, for the values: $p = 0.04, 0.4, 4, 40 [ph/sec/cm^2]$, from top to bottom. The $p = 0.4[ph/sec/cm^2]$ line is our detection threshold.
The number density is uniform in the rescaled plane as expected and it drops to zero below the detection threshold. 
\label{fig:Lzplot}
\end{figure}

To illustrate the distribution of bursts, we display the bursts in the $L-z \ (Luminosity - Redshift)$ plane and in a rescaled plane in which the number of detected bursts (with or without redshift measurement) is proportional to the area (Fig. \ref{fig:Lzplot}). The rather uniform distribution on the rescaled plane is a visual demonstration of the validity of the assumption.

\subsection{A Comparison With the Complete {\it Swift} Sample}
\label{sec:sampleselectconsistency}
Our results are based only on the absorption and photometric determined redshifts of the {\it Swift} sample.
Recently, \cite{Fynbo(2009)} obtained emission lines redshift measurements for GRBs host galaxies that were not measured before. 
Most of those are at low redshifts.
The growing number of emission lines redshifts in the range $0 < z < 1$ raises the question whether our model - based on a sample of redshifts measured from the afterglows - is consistent.
We calculate, using our model, the expected redshift distribution of the entire bursts population and compare it to the number of observed bursts with a known redshift (measured using all methods). Clearly  for any range of redshifts the number of known bursts with a given redshift range should not exceed the number predicted by the model.
\begin{figure}
\includegraphics[width=470pt]{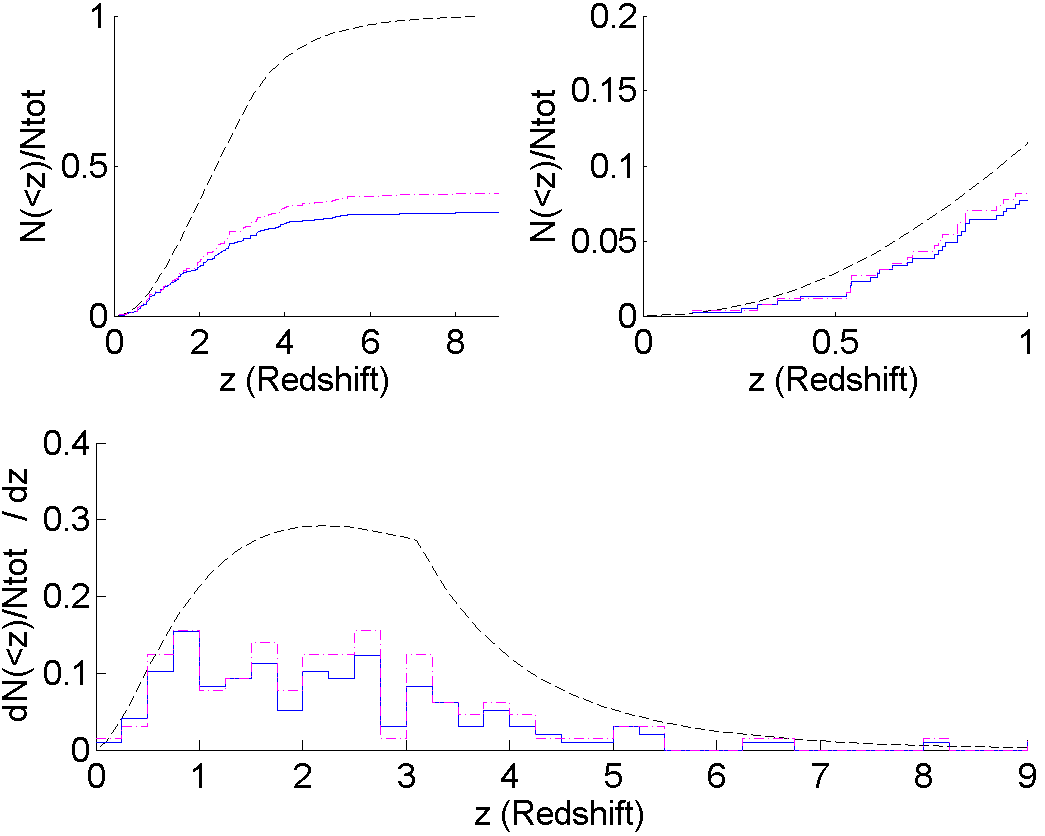}
\caption{Redshifts obtained by any method compared with the total number of bursts predicted by the model}
Upper frames: Left: The cumulative total fraction of bursts predicted by the model (dashed curve) and the detected cumulative fraction of bursts with redshift obtained by any method (solid steps) as a function of the redshift.
Right: Zoom in on the low redshift part.
Lower frame: The number density (per unit redshift) of predicted bursts (dashed curve) and an histogram of bursts with redshift obtained by any method (solid steps). Both are normalized by the total number of bursts.
For all frames: The number of bursts with redshifts which had an at slew, i.e. the time elapsed form the trigger until the first optical observation is less than 300 sec (dashed dotted steps).
\label{fig:NzvsNtot}
\end{figure}
Fig \ref{fig:NzvsNtot} depicts this comparison for the cumulative number and for the counts number in redshift bins. Our model is consistent for all redshifts $\gtrsim 0.4$. However there is an excess of low redshift bursts not predicted by our model. The discrepancy arise due to three bursts with $z < 0.1$ while less than one burst is predicted by the model. The redshift of the three bursts was determined using emission lines but no redshift was detected in this range using absorption lines. These three bursts yield a rate that is significantly higher than predicted by the model. This cannot be explained by a misidentification of the host galaxies because the probability for such an effect is too small \citep{Cobb(2008)}.
These low redshift bursts have low luminosities and they could not have been detected at a much higher redshift. We conclude that they possibly represent a low luminosity population that is different from the majority of the bursts (see a discussion in \S\ref{sec:lowlumpop}).

\subsection{The Redshift Distribution and Expectations for Future Missions}
\label{sec:futmis}
\begin{figure}
\includegraphics[width=470pt]{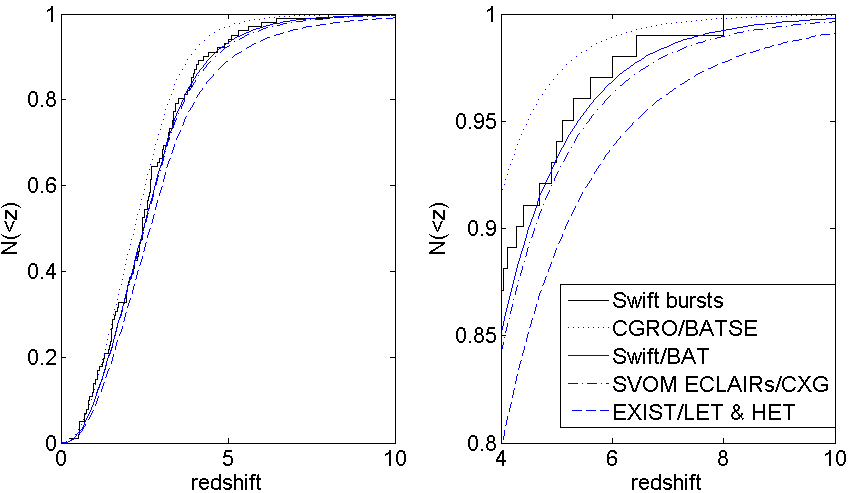}
\caption{The cumulative redshift distribution for {\it Swift} bursts and predictions for other missions}
The cumulative redshift distribution of {\it Swift} observed bursts and the model prediction (solid line).
The KS-test gives a probability of $62\%$. Also presented are the predictions for some past and future missions: BATSE, SVOM and EXIST. Right frame is a zoom-in to the upper right part of the left frame.
\label{fig:cumsum_redshift}
\end{figure}

A particularly interesting question is how many high redshift bursts are expected to be observed. Such bursts are of great interest as they may shed new light on the very early universe. Already now GRB090423 is amongst the most distant and hence the earliest objects observed so far.
Figure \ref{fig:cumsum_redshift} depicts the observed cumulative redshift distribution and the predicted one, for high-redshift bursts for {\it Swift} and for past and future missions: BATSE, EXIST \citep{Band(2008)} and SVOM \citep{Schanne(2008),Gotz(2009)}.
The fraction of bursts with $z > 7, 8, 9, 10, 15, 20$ respectively is shown in Table \ref{tbl:redshift_frac}.
   
Detection of high redshift bursts is one of the main objectives of EXIST. The number of high redshift bursts expected to be detected by EXIST is presented in Table \ref{tbl:redshift_count_EXIST}. 
During a five year mission EXIST will detect many high redshift ($z > 10$) ($\approx$ 30) bursts.
In the best-fit model there is  a good probability for EXIST to detect even a $z > 20$ burst during mission. This is, of course, provided that 
such early bursts exist and that the rate at $z \approx 5-8$ can be extrapolated to such  high redshifts.
\begin{table}
\begin{tabular}{lllllllll}
\hline
fraction of $z >$	& 7	& 8	& 9	& 10	& 15	& 20	\\
\hline
CGRO/BATSE	& 0.4\%	& 0.2\%	& 0.1\%	& 0.05\%	& 0.01\%	& 0.002\%	\\
Swift/BAT	& 1.7\%	& 0.9\%	& 0.5\%	& 0.3\%	& 0.03\%	& 0.01\%	\\
SVOM ECLAIRs/CXG	& 2\%	& 1.2\%	& 0.7\%	& 0.4\%	& 0.06\%	& 0.02\%	\\
EXIST/LET \& HET	& 4\%	& 3\%	& 2\%	& 1\%	& 0.2\%	& 0.06\%	\\
\hline
\end{tabular}
 \caption{The high redshift fraction prediction for {\it Swift} and several other missions}
\label{tbl:redshift_frac}
\end{table}
\begin{table}\renewcommand{\arraystretch}{2}\addtolength{\tabcolsep}{-1pt}
\begin{tabular}{lllllllll}
\hline
bursts per year for $z >$	& 7	& 8	& 9	& 10	& 15	& 20	\\
\hline
EXIST & $23^{+121}_{-12}$ & $14^{+96}_{-8}$ & $9^{+81}_{-6}$ & $6^{+64}_{-4}$ & $0.9^{+25.1}_{-0.7}$ & $0.2^{+11.8}_{-0.16}$ \\
SVOM & $2^{+12}_{-1}$ & $1^{+9}_{-0.5}$ & $0.7^{+6.3}_{-0.5}$ & $0.4^{+4.6}_{-0.3}$ & $0.06^{+1.94}_{-0.05}$ & $0.01^{+0.79}_{-0.007}$ \\
\hline
\end{tabular}
\caption{High redshift detection rates prediction for EXIST and for SVOM. On average a redshift is obtained for only a third of the events.}
\label{tbl:redshift_count_EXIST}
\end{table}

\section{The GRB Rate and the SFR}
\label{sec:contosfr}
The location of long GRBs in the star forming regions led to the expectation that GRB follow the SFR. We turn now to examine this hypothesis. Modeling the SFR throughout the measured redshift range ($0 < z < 10$) is a complicated task, involving observations in various wavelengths and various assumptions on observational proxies for the SFR as well as correcting due to obscuration, absorption and selection effects.
In a classical work Madau \citep{Madau(1996),Madau(1998),Madau(1997)} considered the SFR per comoving volume vs redshift and found a rise from present to $z \approx 1-2$, and then a comparable decline to $z \approx 5$. This form of the SFR vs redshift has become known as the ``Madau plot''. A number of developments (e.g., dust corrections, submm results, new estimates of the SFR at low redshift) led to changes in the shape of the SFR.
Following these developments \cite{Rowan-Robinson(1999)} and \cite{Porciani(2001)} suggested that the SFR rises by a factor of 10 - 20 from $z=0$ to $z \simeq 1$. The rate at higher redshifts ($z > 2$) was undecided at that time and models for the SFR at higher redshift included flat as well as rising and declining functions \citep{Porciani(2001)}. Later, \cite{HB(2006)} showed that at high redshift ($z > 4$) the SFR declines. Recently, several papers estimated the SFR using the new HST WFC3/IR camera \citep{Bouwens(2009a),Oesch(2009),Bunker(2009),McLure(2009),Yan(2009),Bouwens(2009b)}. These works suggest a decline in the SFR for $z \gtrsim 4$ up to $z \sim 8-9$ . In the following we use \cite{Bouwens(2009b)}  as representative of these new high-z SFRs. 
Despite all the observational advances, there are still different models of the SFR even at low ($z < 1$) and intermediate ($1 < z < 3$) redshift. On one hand, the widely used \cite{HB(2006)} piecewise linear fit finds a factor 10 rise in the SFR from $z = 0$ to $z = 1$ and an almost constant rate from $z = 1$ to $z=4.5$ (which follows by a steep decline at higher redshifts) .On the other hand \cite{Bouwens(2009b)} use data from \cite{Schiminovich(2005)} and \cite{Reddy(2009)} for $0 < z < 2$, $2<z<3$ respectively. Their data compilation can be fairly modeled ($\chi^2 = 10.8$ for 9 d.o.f.) by a broken power law rising a factor of 20 from $z = 0$ to $z = 3.3$, then declining as $(1+z)^{-5.8}$ for $3.3 < z$. Note that a rise up to $z=2-3$ is also suggested by \cite{HB(2006)} when fitting to a \cite{Cole(2001)} form. This seemingly mild disagreement between different SFR models have a crucial implication when comparing the GRB rate to the SFR.

\begin{figure}
\includegraphics[width=470pt]{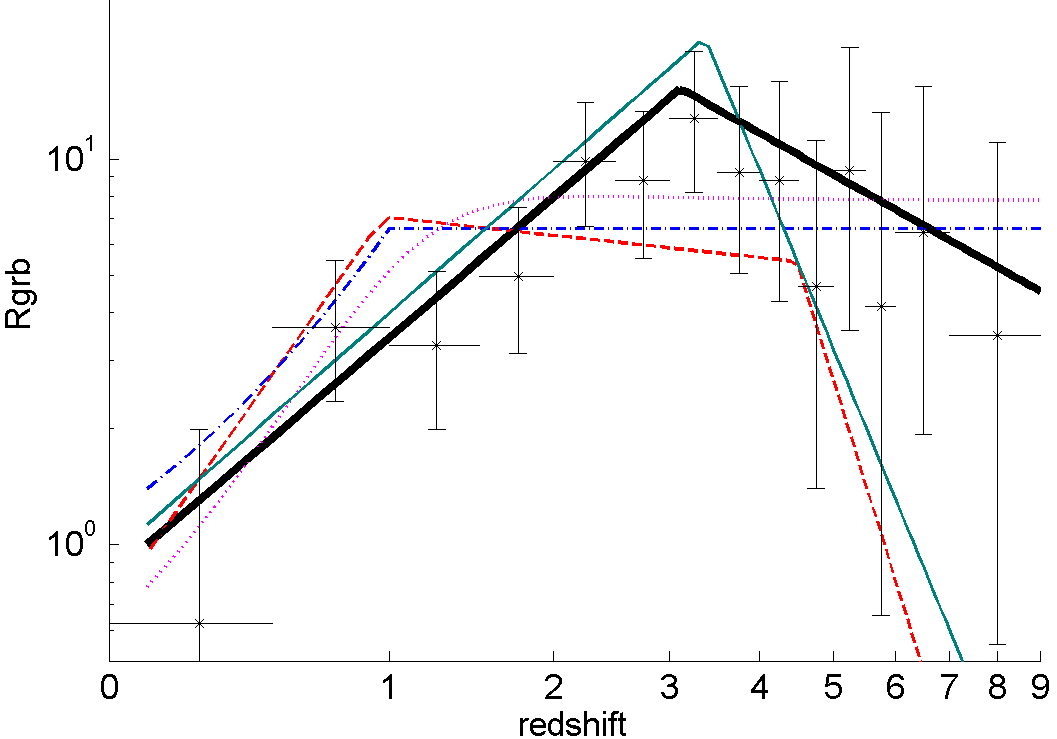}
\caption{GRB event rate and several star formation rates}
The results for the rate, in 1/2 unit binning. Best fit for a broken power law - heavy black solid line. \cite{HB(2006)} SFR - red dashed line, \cite{Bouwens(2009b)} SFR - cyan solid line. SF2 of \cite{Porciani(2001)} - magenta dotted line. \cite{Rowan-Robinson(1999)} SFR - blue dashed dotted line.
\label{fig:SFRsplot}
\end{figure}

The same tests used in \S\ref{sec:consistency}, can be used now to check the consistency of the data with a rate  proportional to the SFR. An obvious problem is that SFR is not uniquely determined and hence we consider four different possible functions: \cite{HB(2006)} (denoted HB), SF2 of \cite{Porciani(2001)} (denoted PM SF2), \cite{Rowan-Robinson(1999)} (denoted R-R) and \cite{Bouwens(2009b)} (denoted B09). Note that HB and B09 decreases at large redshift, while PM SF2 and R-R stay constant. We compare, first, the SFRs to the binned rate we obtained by inverting the data (see Figure \ref{fig:SFRsplot}) and we calculate the $\chi^2$ for both $1/3$ and $1/2$ binning. We find acceptable reduced $\chi^2$ values (see Table \ref{tbl:SFRsstats}) for all the functions. However, a comparison of the overall observed redshift and luminosity distributions with those predicted by models in which the GRB rate is fixed by a given SFR reveals that the 2D K-S or the K-S tests for the peak flux and the redshift distributions show inconsistency for the first three functions (HB, PM SF2, R-R). We find, however, consistency for the last one (B09).

Next, we optimize the \lumf for a given SFR. We take the GRB rate as known following a model of the SFR and obtain the best fit \lumf  by solving equation \ref{eqn:lumf}.
We now perform a 2D K-S test  as well as K-S tests for the peak flux distribution and for the redshift distribution. The results of the statistical tests  are shown in Table \ref{tbl:SFRsstats}. 
Even though the fit improves still the first three SFR models fail the KS tests.
The last SFR model (B09) is, of course,  consistent.

We attribute the consistency of the GRB rate with the B09 SFR model but not with the first three SFR models to the difference between the B09 SFR and the other three models in the range $1 < z < 3$. While B09 keeps increasing in this range the first three are constant. 
This difference is crucial and understanding the SFR at this region is critical for the question whether the GRB rate follows the SFR or not. 
While there are differences at higher redshifts,
the paucity of data  leads to wide error range in that region allowing the GRB observations to be consistent with SFRs that are constant, decreasing or even increasing at large z.
\begin{table}
\begin{tabular}{llllllllll}
\hline
& reduced $\chi^2$	& reduced $\chi^2$	& 2D K-S	& peak flux	& z					& $L*$ & $\alpha$ & $\beta$ & $\rho_0$ \\
& (1/3 bins)				& (1/2 bins)				&					& K-S test	&	K-S test	& & & \\
\hline
\hline
this paper    	& 0.24	& 0.23	& 0.96	& 0.82	& 0.62	& 52.53	& 0.17	& 1.44	& 1.25 \\
R-R(1999)	& 0.69	& 0.85	& $2\cdot 10^{-3}$	& $1\cdot 10^{-3}$	& $9\cdot 10^{-5}$	& 52.53	& 0.19	& 1.44	& 2.04	\\
R-R(1999)	& 	& 	& 0.05	& $2\cdot 10^{-5}$	& 0.01	& 52.46	& -0.06	& 1.48	& 1.40	\\
HB	& 1.14	& 1.38	& $1\cdot 10^{-5}$	& $1\cdot 10^{-4}$	& $4\cdot 10^{-8}$	& 52.53	& 0.19	& 1.44	& 1.30	\\
HB	& 	& 	& $2\cdot 10^{-3}$	& $6\cdot 10^{-9}$	& $6\cdot 10^{-5}$	& 52.49	& -0.12	& 1.46	& 0.83	\\
PM SF2	& 0.53	& 0.58	& 0.05	& 0.05	& 0.01	& 52.53	& 0.19	& 1.44	& 1.01	\\
PM SF2	& 	& 	& 0.17	& 0.01	& 0.07	& 52.47	& -0.00	& 1.48	& 0.75	\\
\bf B09 	& \bf 0.57	& \bf 0.60	& \bf 0.21	& \bf 0.93	& \bf 0.14	& \bf 52.53	& \bf 0.19	& \bf 1.44	& \bf 1.22	\\
\bf B09	& 	& 	& \bf 0.15	& \bf 0.65	& \bf 0.19	& \bf 52.53	& \bf 0.08	& \bf 1.47	& \bf 0.99	\\
\hline
\bf Butler et al.(2009)	& \bf 0.27	& \bf 0.27	& \bf 0.76	& \bf 0.65	& \bf 0.71	& \bf 52.53	& \bf 0.19	& \bf 1.44	& \bf 0.80	\\
\bf Butler et al.(2009)	& 	& 	& \bf 0.66	& \bf 0.46	& \bf 0.50	& \bf 52.47	& \bf 0.00	& \bf 1.47	& \bf 0.58	\\
\hline
\end{tabular}
\caption{Statistical tests for our models and for models following one of the SFRs considered (upper part) and for GRB rate from other studies (lower part). The consistent models are marked with a bold font. For each model we show the test with \lumf from our results (first line) and with \lumf that best fit observations after forcing the rate to follow the SFR (second line).}
\label{tbl:SFRsstats}
\end{table}

The comparison Table shows that the \lumf parameters depends very weakly on the GRB rate model. $\alpha$ and $\rho_0$ changes within their 95\% error range, while $L*$ and $\beta$ stay almost unchanged. This illustrates the power of the method and the robustness of the \lumf results, in particular for $L*$ and $\beta$.

\subsection{Comparison with other works}
Our results are in agreement with most previous  works on the GRB rate. 
Our results agree with \cite{Daigne(2006)} who compared the {\it Swift} data with \cite{Porciani(2001)} SFR models and concluded that the rising model SF3 is the only one consistent (Our model is also consistent with SF3). We stress however, that the recent work on the SFR suggests that it decreases at high redshift and thus  SF3 is, most likely,  not a viable SFR model. Our results are also consistent with \cite{Guetta(2007)} who find s that at $z>2.5$ the GRB rate is significantly higher than the R-R or PM SF2 SFR. 

A comprehensive work studying the \lumf and the rate of long GRBs was recently published \citep{Butler(2009)}. While this work uses a different sample, adopts somewhat different assumptions and uses other methods, a comparison of the results can be very useful. We find a good agreement for the low luminosity power law index $\alpha = 0.22^{+0.18}_{-0.31}  (our \: results: 0.2^{+0.2}_{-0.1})$ and for the break luminosity $L* = 10^{52.74 \pm 0.43} (our \: results: 10^{52.5 \pm 0.2})$. However we find different high luminosity power law index $\beta = 2.9^{+2.1}_{-1.1} , (our \: results: 1.4^{+0.3}_{-0.6})$. The differences may be explained by the fact that \cite{Butler(2009)} use a luminosity which is a time averaged whereas our luminosity is the peak luminosity. We find a nice agreement between the models for the GRB rate. \cite{Butler(2009)} find a rising rate for $0 < z < 4$ - first at slope $3.1 \pm 0.7$ for $0 < z < 1$ and later at slope $1.4 \pm 0.6$ for $1 < z < 4$. This is not very different from our results, recalling that the break at $z = 1$ was not a free parameter in their model. The decline slope for $z > 4$ is $-2.9^{+1.6}_{-2.4}$, but with the big uncertainties it also matches our model. It is thus not surprising that their model show consistency with the statistical tests (see Table \ref{tbl:SFRsstats}).
Another useful comparison is with \cite{Kistler(2009)} who modeled the bias of the GRB rate with SFR in the range $0 < z < 4$ and used it together with the high-z bursts data to estimate the high-z SFR. The high-z GRB rate they found is roughly constant, in agreement with our results.

\section{Summary and Conclusions}
\label{sec:summary}
We find that  the GRB rate increases  up to redshift $\simeq 3$ and it decreases at $z > 3$ ($n_2 \simeq -1.4$), with 68\% confidence limits ranging from a steep decline ($n_2 \simeq -2.4$), to a positive incline ($n_2 \simeq 1$). The rate is  compatible, of course,  with a constant rate at higher redshifts.
The model is consistent with all statistical tests and thus we can accept the basic assumption that the luminosity function dose not evolve with time.

\subsection{GRB Rate and the SFR}
The comparison between the SFR and the GRB rate seems to be inconclusive. This arises because  of the differences between different estimates of the  SFR at the intermediate redshift range $1 < z < 3$. The rate we find is consistent with \cite{Bouwens(2009b)} (B09) \citep[that follows][]{Schiminovich(2005)} that describes a rising SFR from present up to $z \approx 3$. It is inconsistent with other SFRs \citep{Rowan-Robinson(1999),Porciani(2001),HB(2006)} that suggest a constant rate at this regime.

At high redshifts the GRB data is sparse. The best fit result decreases slower than the most recent B09 SFR (or even the HB SFR) suggesting possibly higher GRB rate. However fast decrease, like B09, cannot be ruled out while a flat or even slowly increasing rate are also consistent.
A larger GRB rate at high redshift (compared to the SFR) can be explained within the framework of the massive stellar collapse model due to metallicity. \cite{Woosley(2006)} suggests that GRB rate follow the low-metallicity part of the star formation.
We cautiously note however that \cite{Fynbo(2009)} recently found that the optical afterglow spectroscopy sample is biased against measuring redshift at high metallicity environments, meaning that the result might be an artifact of a selection effect.
Another clue can be taken from \cite{Fruchter(2006)} who found that the GRBs distribution within the host galaxies dose not follow the light distribution but rather some power ($>1$) of the light distribution, i.e. higher GRB density in the denser star forming regions. This is in contrast to the core-collapse supernovae distribution that follows the light distribution, indicating the GRB progenitors are different from SN progenitors and hence their rate might be different.

\subsection{Low Luminosity Bursts}
\label{sec:lowlumpop}
We find an overall consistency when comparing our model and the full sample of GRBs with redshift. However, we also find  three low redshift, low luminosity bursts (with emission lines redshifts) that are not expected by the model prediction. The faintest burst in our sample GRB050724 has a luminosity $L_1 = 10^{50.4} [erg/sec]$.
By applying our best fit model, we expect 0.9 bursts with luminosity $L \leq L_1$, in the time span of our sample (4.5 years).
{\it Swift}s weakest burst, GRB060218 \citep{Cusumano(2006)} with luminosity $L_2 = 10^{47.4} [erg/sec]$,
is not in our sample because it has only emission lines redshift. 
Assuming that we can extrapolate the low end of our luminosity function we expect $2.5 \cdot 10^{-6}$ bursts with $L \leq L_2$.
Even when applying the 95\% level of the parameter $\alpha =0.38$, we expect no more than $5 \cdot 10^{-5}$ bursts with $L \leq L_2$.
This implies that this burst represents a population of fainter GRBs, with much higher event rate, which cannot be directly extrapolated from the stronger GRB population \citep[in agreement with e.g.][]{Soderberg(2006),Cobb(2006),GuettaDV(2007),Liang(2007)}.
The emission lines redshifts sample includes two other low luminosity bursts: GRB051109 and GRB060505 with $z = 0.08, 0.089$ and $L = 10^{48.6} [erg/sec], 10^{49.4} [erg/sec]$.
The detection probability of such bursts, according to our model is $<2 \cdot 10^{-3}, <2 \cdot 10^{-2}$ respectively. These three low luminosity bursts must belong to a different and a distinct group and we remove them from the analysis when checking for consistency in \S\ref{sec:sampleselectconsistency}.

\subsection{High Redshift Bursts}
Higher redshift bursts are most interesting as they can provide clues on the very early universe.
Extrapolating the rate for very high redshifts, we expect $\sim 0.9$ bursts with $z > 8$ and $\sim0.5$ bursts with $z > 9$ (bursts with measured redshift), detected by {\it Swift} in the time span of our sample (4.5 years). This is consistent with the observation of one $z > 8$ burst. Future missions like EXIST can find many more such bursts, even dozens of $z > 10$ bursts (see \S\ref{sec:futmis}) provided that a simple extrapolation indeed hold to such high redshifts.

\subsection{The Local Event Rate}
The local event rate found is $\rho_0 = 1.3^{+0.6}_{-0.7} [Gpc^{-3} yr^{-1}]$,  for bursts with $L \geq 10^{50} erg/sec$ . With one galaxy in $100 Mpc^3$ this rate is equivalent to 1 event per galaxy per $10^7$ years! Taking into consideration the beaming factor of about 50 \citep[see][]{Guetta(2005)}, the total events rate is about 1 event per galaxy per $2 \cdot 10^5$ years.  This should be typical to our galaxy as a recent estimation to the SFR in our galaxy \citep{Robitaille(2010)} finds $\dot{\rho} = 0.68 - 1.45 M_{\odot} /yr$, which is equivalent to the SFR in the local universe for a galaxy in a volume of $100 Mpc^3$. One implication of this result will be on the possible association of GRBs with global extinctions of biological spices. These occurred on Earth a factor of 10 times less frequent, about once every 100 Myr.
These two rates suggest that a typical Galactic GRB pointing to Earth does not cause a major extinction event. 

\section*{Acknowledgments}
We thank Shiho Kobayashi, Ehud Nakar and Elena Rossi for fruitful discussions.
The research was supported by an ERC Advanced Research Grant, by the ISF center for High Energy Astrophysics and by the Israel-France program in Astrophysics grant.

\appendix
\numberwithin{equation}{section}
\section[A]{The luminosity}
\label{sec:lumcalc}
In this work we used only data collected by {\it Swift}. The peak-flux of each burst, measured in {\it Swift}'s BAT detectors band 15keV - 150keV \citep{Barthelmy(2005)}. For most of the bursts we cannot estimate the luminosity in the full $\gamma$-range (1keV - 10MeV) as the spectral shape (the Band-function) \citep{Band(1993)} is poorly known. We still need however to have some quantitative measure of the luminosity that will be defined uniformly for all burst with a known redshift. To do so we use an average characteristic Band function, $E_{peak}=511keV$ (in source frame), $\alpha=-1$, $\beta=-2.25$ \citep{Preece(2000),Porciani(2001)} to estimate luminosity, 
using the same parameters for all bursts, thus the estimated luminosity is proportional to the measured $peak \ flux$ (p):
\begin{equation}
 L_{iso} = p 4 \pi D(z)^2 (1 + z) k(z) C_{det} \ ,
\end{equation}
$D(z)$ is the proper distance at redshift $z$, $C_{det}^{-1}$ is the fraction of the total $\gamma-ray$ luminosity detected in the detectors energy band for source at redshift = 0,
\begin{equation}
C_{det} = \frac{\int_{1keV}^{10MeV}E N(E) dE }{\int_{E_{min}}^{E_{max}} N(E) dE} \ ,
\end{equation}
$k(z)$ is the k-correction for the given spectrum at redshift z,
\begin{equation}
k(z) = \frac{\int_{E_{min}}^{E_{max}} N(E) dE}{\int_{(1+z)E_{min}}^{(1+z)E_{max}} N(E) dE} \ , 
\end{equation}
where $N(E)$ is the Band function, $E_{min} = 15 keV$, $E_{min} = 150 keV$.
The values in this paper are the luminosity in the range [1keV - 10MeV].

\begin{figure}
\includegraphics[width=391pt]{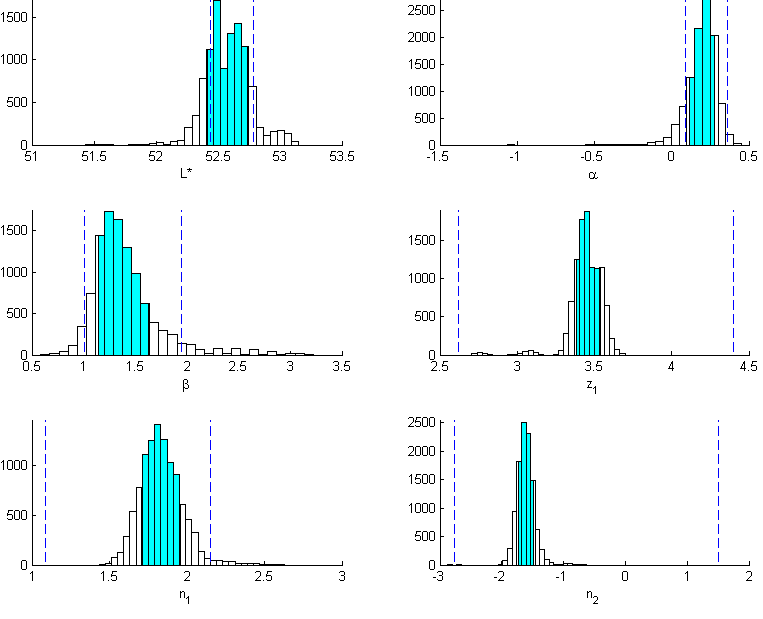}
\caption{Distributions of our fits parameters for the \lumf ($L*,\alpha,\beta$) and for the rate ($z_1,n_1,n_2$)  when choosing the $\gamma$-ray spectra (Band function) parameters randomly from their distributions. The shaded area contains the central 68\% of the distribution, the dashed lines are the 68\% confidence range of our reported results.}
\label{fig:vary_bandf}
\end{figure}

 To study the dependence of our results on this approximation of an average Band function, we have preformed a simulation where the spectrum of the bursts is not universal but drawn randomly from the known distribution \citep{Preece(2000)}. Repeating the analysis in the paper 1000 times, each time calculating the luminosity of a burst using a band function with parameters randomly drawn - independently for each burst - from the distribution. Fig \ref{fig:vary_bandf} show the simulation results as an histogram for each of the parameters of the \lumf and the rate fits. All the distributions concentrate within the $68\%$ error range of the original results, demonstrating the robustness of the results and its insensitivity to the details of the distribution of spectral parameters. 

\section[B]{The Redshift Detection Probability}
\label{sec:measureprob}
We examine now the redshift detection probability as a function of the $peak \ flux$.
We consider here two effects: first the probability to detect the GRB and second the probability to measure the redshift, for a given detected burst.

The probabilities to detect a GRB, or to measure a redshift, are a function of the burst energy, its duration and other factors, as described by \citep{Band(2006)}. We restrict ourselves here only to the dependence of these probabilities in the $peak \ flux$, since this is the quantity we use for the analysis in this paper.

\subsection{GRB Detection Probability}
\begin{figure}
\includegraphics[width=391pt]{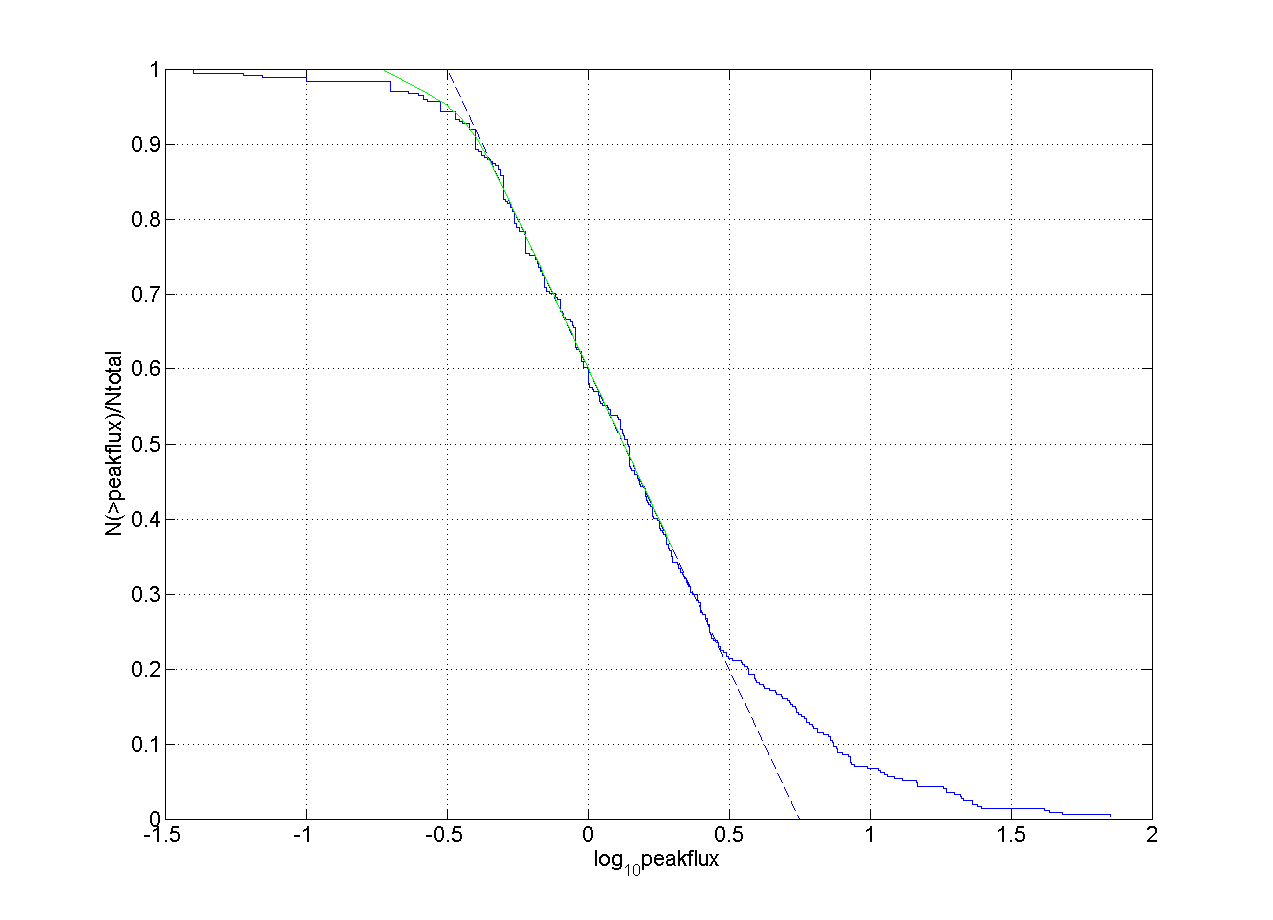}
\caption{The accumulated number of bursts as function of log(peak flux)}
\label{fig:bursts_count}
\end{figure}

The simplest model often used is of a sharp threshold: No detection with $p < p_{lim}$, but 100\% detection of bursts with $p \geq p_{lim}$. 
For this model the value used at {\it Swift}'s main detection band $[15 - 150]keV$ is $p_{lim} = 0.4 ph/cm^2/sec$. (see \cite{Guetta(2007)}, \cite{Gorosabel(2004)}). 
In the plot of the accumulated number of bursts as a function of $log(p)$, a linear relation appears for fluxes that are low or medium but above the threshold mentioned above.
Although we do not try to give a theoretical explanation for this result, we do not expect, however any strong deviation from that connection for lower fluxes, since our models predict a slow gradual smooth change in $dN / dlog(p)$ and we can adopt this values at least as an order of magnitude estimators, to yield a continues threshold estimation.
Assuming that the deviation from linearity for fluxes below $p_{lim}$ is only due to a lowered detection sensitivity, we can extrapolate the predicted number of bursts for lower fluxes and extract the detection sensitivity by comparing the number of detected bursts to the predicted number.
Figure \ref{fig:bursts_count} shows this: accumulated number of bursts vs. log peak flux, the linear fit and our fit.
\begin{equation}
\theta_{\gamma}(p) =
\left\{
\begin{array}{ll}
 \frac{(1+c)}{2}+ \frac{(1-c)}{2}erf(d \cdot log(p/p_0) & 0.2 \leq p \ , \\
 0 & p < 0.2 \ , 
\end{array}
\right. 
\end{equation}
with the parameters found:
\begin{equation*}
 c = 0.25, d = 10, log_{10}p_0 = -0.42 \ .
\end{equation*}
\subsubsection{Redshift Measuring Probability}
Redshifts are measured only for a moderate fraction of the detected bursts. A few (5-10\%) are too weak, some don't have a clear redshift signature and some are not measured because of lack of observational resources.

\begin{figure}
\includegraphics[width=391pt]{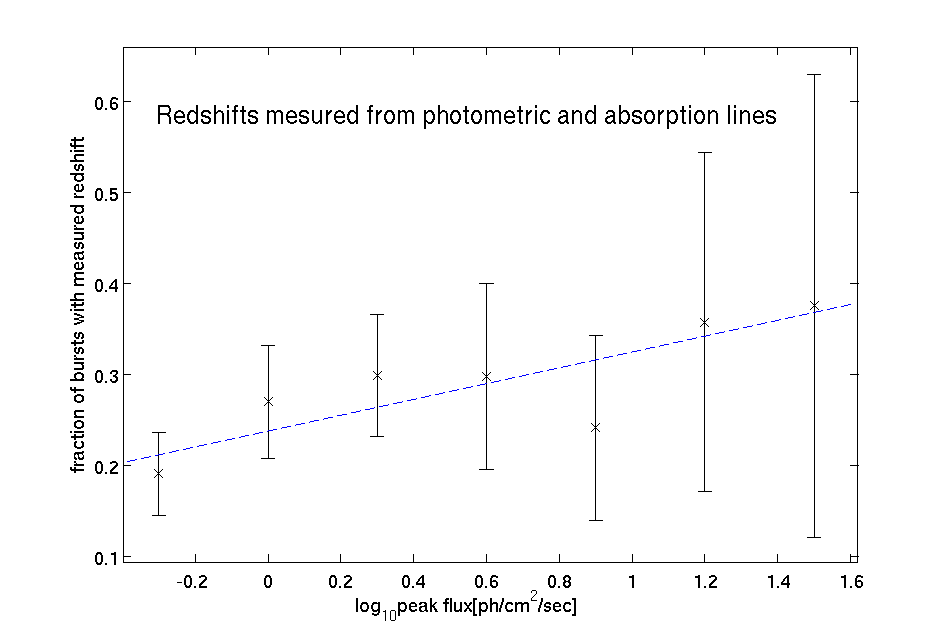}
\caption{The fraction of bursts with a measured absorption lines and photometry redshift, as function of log(p)}
\label{fig:zdetprob_ap}
\end{figure}

When we consider the fraction of redshift-measured GRBs to the total number of GRBs detected, we see a trend of increase in this fraction with the measured peak number flux of photons in the detectors main band: $15keV - 150keV$, p (see figure \ref{fig:zdetprob_ap}). We used a linear regression to approximate the relation: 
\begin{equation} \theta_{z}(p) / \theta_{\gamma}(p) = a \cdot log(p) + b \ , \end{equation}
Where , $\theta_{\gamma}(p)$ is the detection probability and $\theta_{z}(p)$ is the probability that a redshift will be measured. The parameters found in fit are: $a = 0.080\pm 0.078$ and $b = 0.242\pm 0.054$. with $\chi^2 = 1.01$ at 5 degrees of freedom, giving a rejection probability of 0.038 .
We expect the fraction of number of bursts with redshift $N_{z}(p)$ to the overall number of bursts $N_{\gamma}(p)$, for any given $p$, to obey the relation:
\begin{equation}
 \frac{\theta_{z}(p)}{\theta_{\gamma}(p)} = \frac{N_{z}(p)}{N_{\gamma}(p)} \ .
\end{equation}

Figure \ref{fig:zdetprob_ap}. depicts the fraction of bursts with a measured redshift and the linear fit, for the absorption and photometry redshifts.
Although the fit is acceptable the significant of the effect is just one standard deviation $(\sigma)$ - so with the current observations the option of no dependence of redshift detection probability with peak flux $(a = 0)$ is still marginally consistent.

\begin{figure}
\includegraphics[width=391pt]{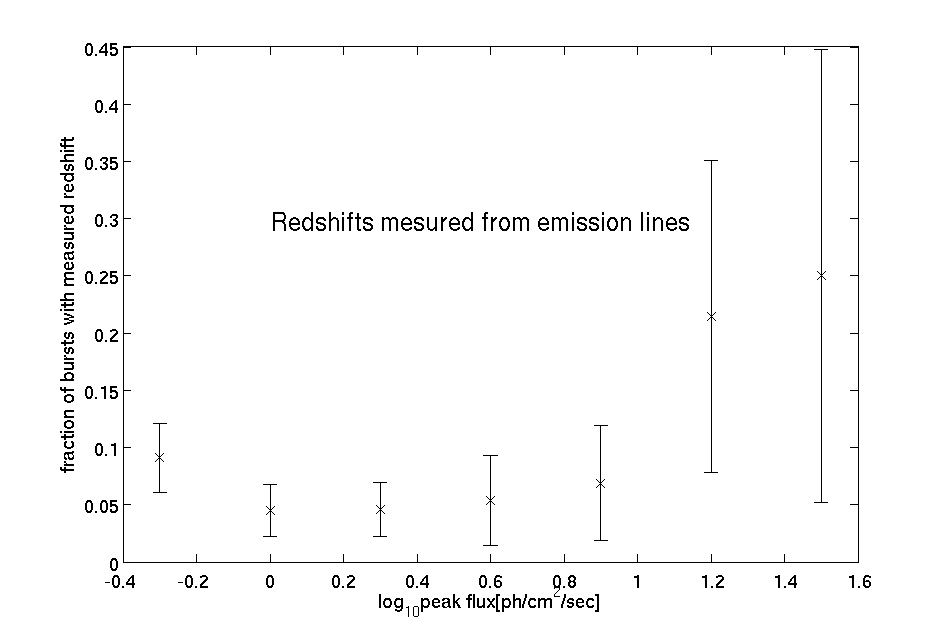}
\caption{The fraction of bursts with a measured emission lines redshift, as function of log(p)}
\label{fig:zdetprob_eh}
\end{figure}

Whereas the detection fraction can be fitted well with a linear model for the redshifts obtained using absorption lines, the case is different when considering redshifts obtained using emission lines.
Figure \ref{fig:zdetprob_eh}. shows the measured redshift fraction and the linear fit for the emission lines redshifts.
The emission lines redshifts are obtained preferably for high flux bursts, but very few are obtained for low and medium fluxes: $24\%$ for $log_{10}p > 1$ and only $6.5\%$ for $log_{10}p \leq 1$.
This feature of the emission lines redshifts is associated with a strong bias toward lower redshifts as shown on Figure \ref{fig:zmethodsdist}. These results supports our approach of selecting only the absorption lines and photometry redshifts as it make a sample which is less biased and easier to model.

\begin{figure}
\includegraphics[width=350pt]{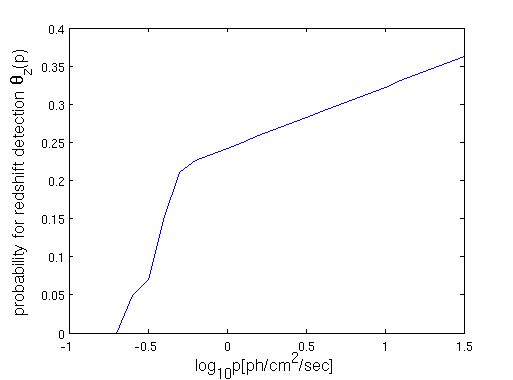}
\caption{the probability for measuring redshift by absorption lines or by photometry $\theta_z$, as a function of log(peak flux)}
\label{fig:theta_z}
\end{figure}

The product of the above probabilities for burst detection and redshift measurement, gives the total probability that we detect a burst and measure its redshift $\theta_z$.
Figure \ref{fig:theta_z}, shows the function $\theta_z$, which have the form:
\begin{equation}
\theta_z(p) =
\left\{
\begin{array}{ll}
 (a log(p) + b)(\frac{(1+c)}{2}+ \frac{(1-c)}{2}erf(d \cdot log(p/p_0)) & 0.2 \leq p \ , \\
 0 & p < 0.2 \ ,  
\end{array}
\right. 
\end{equation}
with
\begin{equation*}
 a = 0.08 ,b = 0.242, c = 0.25, d = 10, log_{10}p_0 = -0.42 \ .
\end{equation*}

To check the effects of the burst detection probability and the redshift measurement probability, we repeated the process described in this paper, with all four options of taking or not taking into account each of the modified probabilities.
The best fit results for the parameters 
are shown in Table \ref{tbl:detprob_effects}.
\begin{table}
\begin{tabular}{ll|lll|lll}
\hline
Prob. $z$ & Prob. $\gamma$ & $L*$ & $\alpha$ & $\beta$ & $z_1$ & $n_1$ & $n_2$\\
\hline
-	& -	& 52.54	& 0.17	& 1.32	& 3.13	& 1.85	& -1.38 \\
-	& +	& 52.53	& 0.04	& 1.32	& 2.25	& 1.53	& -0.00 \\
+	& -	& 52.53	& 0.30	& 1.44	& 3.08	& 2.28	& -1.06 \\
+	& +	& 52.53	& 0.17	& 1.44	& 3.11	& 2.07	& -1.36 \\
\hline
\end{tabular}
 \caption{Model's best-fit parameters when taking (+) and when not taking (-) each effect into account.}
\label{tbl:detprob_effects}
\end{table}
For both corrections, we find a non-negligible effect on the results, although the deviations induced on the parameters are in most cases within the statistical error ranges of the analysis.
When taking both effects into account, the models give somewhat better results in the various statistical tests. Therefor, the function $\theta_{z}(p)$ that take both effects into account is used in this work.
\end{document}